\documentclass[a4paper,11pt]{article}
\pdfoutput=1 

\usepackage{jheppub} 

\usepackage[T1]{fontenc} 

\title{\boldmath Now, and the Flow of Time}

\author[a]{Richard A. Muller,}
\author[b,c]{Shaun Maguire}

\affiliation[a]{Department of Physics, University of California, Berkeley, California 94720, USA}
\affiliation[b]{Institute for Quantum Information \& Matter and Walter Burke Institute for Theoretical
Physics, California Institute of Technology, Pasadena, California 91125, USA}
\affiliation[c]{Department of Mathematics, California Institute of Technology, Pasadena, California 91125, USA}

\emailAdd{ramuller@lbl.gov}
\emailAdd{smaguire@caltech.edu}

\abstract{The progression of time can be understood by assuming that the Hubble expansion takes place in 4 dimensions rather than in 3. The flow of time consists of the continuous creation of new moments, new \textit{nows}, that accompany the creation of new space. This model suggests a modification to the metric tensor of the vacuum that leads to testable consequences. Two cosmological tests are proposed, but they present both experimental and theoretical problems. A more practical and immediate test is based on a predicted lag in the emergence of gravitational radiation when two black holes merge. In such mergers (as recently observed by the LIGO team), a macroscopic volume (millions of cubic kilometers) of space is created in the region in which the gravitational wave is generated; this one-time creation of new space should be accompanied by the creation of detectable level of new time, resulting in a time delay that could be observed as a growing lag in the emission of the wave as the merger takes place.}

\keywords{LIGO, gravitational waves, arrow of time} 

\begin{document} 
\maketitle
\flushbottom

\section{Introduction}
\label{sec:intro}
Time is unified with space through the Minkowski concept of space-time, yet time and space have qualitatively different behavior in a way that goes beyond a minus sign in the metric. Given any coordinate system, we can stand still in space but not in time; time inexorably flows. The rate of flow depends on the velocity of the local Lorentz frame and on the gravitational potential.  Yet this description of the relative changes in the rate of flow does not address the key disparity that time flows yet space doesn't. 

In many ways a simpler problem is the arrow of time, the intriguing question of why time flows forward rather than backward, given that most of the fundamental equations of physics show a forward/backward symmetry. Eddington \cite{Eddington} attributed the arrow of time to the second law of thermodynamics, the statement that the entropy of the universe always increases, and that this is the only ``law'' of physics that contradicts time-reversal symmetry (or, at least, it was at the time Eddington conceived the theory). The recent discovery of time-symmetry violation in B decay \cite{Bmeson} suggests that the direction of time might be set by something more fundamental. 

The Eddington proposal, that the arrow of time is related to entropy increase, has many shortcomings.  At its heart, the second law is basically tautological; it consists of the statement that the future will probably be composed of states that, because they have high multiplicity, are more likely.  Arguably, the physics of the second Law is found in ergodic hypothesis, the assumption that all accessible states are equally likely so ones with high multiplicity are more probable. But that principle bears little relationship to the flow of time.

When Eddington proposed the concept, he was unaware of the fact that there is substantially more entropy in the cosmic microwave radiation than in all of the visible matter of the universe, by a factor of about 10 million. Moreover, because that radiation expands adiabatically with the Hubble expansion, its entropy is not changing.  In addition, it is widely thought that there is even a vaster store of entropy on the surfaces of massive black holes, and perhaps even more on the event horizon of the universe. The entropy of these regions is thought to be increasing, but they are so remote from the earth (signals from these surfaces cannot reach us in finite time), that it is hard to understand why they should have an effect on our local time. In the Eddington theory, the arrow of time is set remotely and universally, with no correlation expected between local variations in the rates of entropy and of time. Contrast this to the general theory of relativity, which correctly predicted that local gravitational potential has an immediate and (these days, easily) observed effect on local time.  Moreover, the entropy of the Earth is decreasing as it sheds entropy to infinity; it is likely the entropy of the Sun (not including the radiation which it has discarded) is decreasing. This leads to the result that the entropy of all known matter in the universe, with the exclusion of photons lost to space, is decreasing. 

And finally, unlike the general theory of relativity, the arrow of time / entropy speculation leads to no testable predictions that could falsify it.  By the standards of Karl Popper, it does not rank as a valid physics \textit{theory}. Not only can it not be falsified, but it can not even be verified--unlike string theory which, although not falsifiable with current predictions, at least does predict possibly observable particles, extra compact dimensions that could be detected (but haven't been so far), and subtle correlations in the cosmic microwave background \cite{CCP}.

In this paper we explore a possible cosmological origin for the flow of time. The basic thesis was previously discussed by one of us in the book \textit{Now}\cite{Muller}. It is motivated by the standard Friedman-Lemaître-Robertson-Walker (FLRW) approach in which the universe is modeled by a homogeneous and isotropic distribution of galaxies with fixed coordinates, and the Hubble expansion is described by a changing metric.  In the FLRW metric, new space is being continually created between the galaxies, and that is what gives rise to the observed redshifts; the galaxies are not \textit{moving} (at least in the FLRW coordinate system), but the space between them is increasing as the metric changes.  We postulate that the increase in space is accompanied by an increase in time, by the creation of new moments of time. Unlike the picture drawn in the classic Minkowski space-time diagram, the future does not yet exist; we are not moving into the future, but the future is being constantly created. 

The moment we experience as ``now'' can be identified as the instant of new time that was just created. Although our model is beginning to sound like philosophy \cite{McTaggart} rather than physics, we believe that this model of time leads to predictions that could falsify or verify the theory.  Irregularities in the flow of time can be detected in astronomy since observations of distant objects show us behavior that took place in the past; if time had slowed or accelerated, we could in principle observe such behavior by looking at distant objects. Such observations are analogous to those of the gravitational redshift, which (according to the equivalence principle) is indistinguishable from that of an accelerating reference frame. The expansion of the universe is slowed by mutual gravitational attraction and accelerated by dark energy; if time is similarly showed and accelerated, there would be an additional non-FLRW frequency shift in observed spectra of distant galaxies. It is not clear, however, that such a shift could be separated from the observed cosmological redshift; such a shift could simply be absorbed into the standard FLRW metric and interpreted as additional dark energy rather than as a change in the rate of time. To separate the effects, we might require a new method for estimating recessional velocity, one that does not depend on redshift. Such an observation might consist, for example, of a changing angular size of a stable object; however, no such observations appear practical for the foreseeable future. 

The inflationary era offers another period when the creation of space accelerated. If the creation of time also accelerated during this era, there may be observable consequences, provided that we some day detect the gravitational waves that we believe were emitted during that era. There is hope that such waves could be observed by their effect on the polarization of the cosmic microwave background. 

Although the inspiration of the theory was found in the expansion of the universe, the implications go beyond cosmology.  In particular, we expect that any change in the amount of local space will result in a corresponding change in the local time.  Put another way, any local creation of new space (from an increase in the spatial components of the metric tensor) will be accompanied by the creation of new local time (an increase in the value of the time component of the metric tensor). 

Note that in standard relativity theory, changes in space intervals are typically accompanied by changes in time intervals. Thus space contraction goes hand in hand with time dilation. We are postulating an additional effect -- that a one-time increase in local space results in a one-time increase in the local time.  Inclusion of a term that accomplishes this might not require any change in the general theory of relativity, but could possibly be accomplished by a change in the assumed energy-momentum tensor of the vacuum, or possibly even in the definition of the tensor for matter.

\section{Experimental Test Using LIGO}
\label{sec:LIGO}

The discovery of two (so far) extraordinary gravitational wave events by the LIGO project gives us optimism that local steps in time could be observed by the resulting lags they would produce. 

The LIGO detections are arguably the first time that we have observed events that involve strong-field (high curvature) gravity; moreover, they probe these strong fields undergoing rapid changes.  From such events we have hope that the \textit{now} theory of new time creation can be verified or falsified, if not with the first two reported events, then in the near future when similar events are discovered with improved signal-to-noise ratio.  

\begin{figure}[htp] 
\centering
\includegraphics[width=5in]{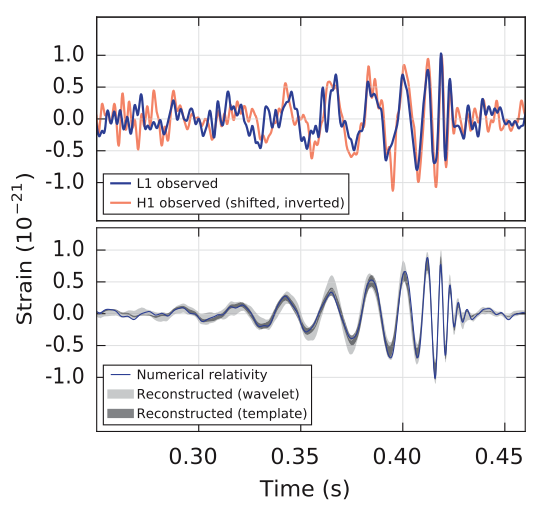} 
\caption{The strain observed in the first reported LIGO gravitational wave, filtered by pre-whitening the signal and then applying a 35-350 Hz band-pass.  From the LIGO paper \cite{LIGO1}.}
\label{fig:firstEvent}
\end{figure}

The first LIGO detection represents the first time that we believe a substantial amount of spatial volume, millions of cubic kilometers, was created in a short and sudden event taking only a few hundredths of a second. This is because the volume surrounding the two original black holes is macroscopically less than the volume in the immediate vicinity of the new black hole resulting from the merger.  According to the \textit{now} theory, this one-time creation of new space should have been accompanied by a one-time creation in new, local time, amounting to about a millisecond.  Moreover that time was created in the region in which the gravitational wave was emitted, and so it could be observed as a lag in the emission of the wave as the merger was taking place.  The reported wave-form of the first detected gravitational wave is shown in Figure \ref{fig:firstEvent}, taken from the LIGO paper \cite{LIGO1}. 

\begin{figure}[htp] 
\centering
\includegraphics[width=5in]{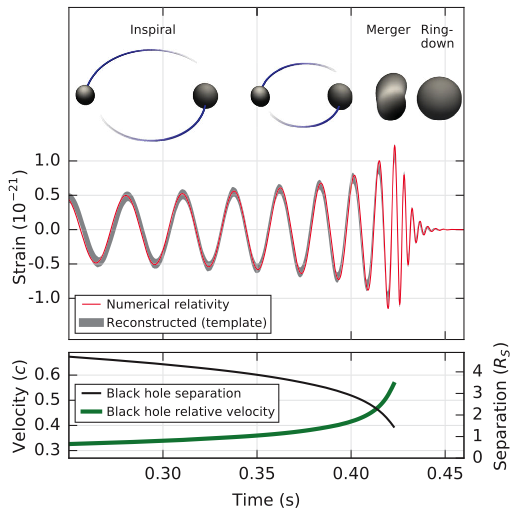} 
\caption{The numerical model for the origin of the LIGO pulse, with no spectral filtering, clearly showing the early 30 Hz oscillations as the two black holes spiraled inward. Taken from the LIGO paper \cite{LIGO1}.}
\label{fig:pulse}
\end{figure}

The shape of the expected pulse, without spectral filtering, is shown in Fig \ref{fig:pulse}, also taken from the LIGO paper \cite{LIGO1}.  This figure also shows a depiction of the merger, and a chart showing the separation and velocity as the merger takes place.  At 0.15 seconds prior to the merger (i.e. at 0.25 sec on the abscissa), the gravitational wave frequency is approximately 30 Hz.  This event has a better signal-to-noise ratio than does the second LIGO event \cite{LIGO2}, and it is the one we will analyze in this paper.

\section{Dimensional Analysis}
\label{sec:DA}

We begin with simple dimensional analysis to estimate the amount of new time that may have been created in the observed LIGO event.  In that event, a black hole with a mass of approximately 62 solar masses was created from two smaller black holes, with 29 and 36 solar masses; about 5\% of the mass energy was converted into gravitational radiation energy. Black holes have a significant effect on space within a few Schwarzschild radii, and it is also within this relatively small region that the observed gravitational radiation was emitted.  In order to get a rough estimate from dimensional analysis, let's assume that an amount of space roughly equal to one Schwarzschild radius in extent was created in this event.  The Schwarzschild radius is given by $R_s = 2GM/c^2=$180 km. With that much space created, and with the unity of space and time, we might expect that the time created to be $t = R_s/c =$ 0.0006 second.  If we include the gravitational redshift, this extra time observed on Earth should appear to be of order 1 millisecond delay in the end of the pulse vs the beginning. This is an encouraging result; although it is too small to be detectable in the first LIGO event shown in Figure 1, it might be seen in future events if they have better signal-to-noise, from being larger or closer to Earth.

\section{Calculation Using the Schwarzschild Metric}
\label{sec:SM}
To make a more precise (although not necessarily more accurate) estimate of the increased space when black holes spiral into each other, we can make a simple estimate of the change in the volume of space near the two original black holes and compare it to the volume of space near the subsequent black hole. For simplicity, we assume static non-rotating Schwarzschild metrics for each of the two merging black holes as well as for the final larger black hole. (The final black hole in the first LIGO event has substantial rotation, but we ignore this.) 

The proper volume from $R_1$ to $R_2$ surrounding a non-rotating Schwarzschild black hole with mass M and Schwarzschild radius $R_s = 2GM/c^2$ is given by \cite{integral}:
\begin{equation}
 V(R_1, R_2, M) =  \int_{R_1}^{R_2} \frac{4\pi r^2}{\sqrt{1-R_s/r  }}dr
\end{equation}
When a black hole is created in previously (nearly) flat space, we take the change of volume created within the radii $R_1$ and $R_2$ to be
\begin{equation}
\Delta V(R_1,R_2,M)  =   V(R_1,R_2,M)-V(R_1,R_2,0)                     
\end{equation}
where the second term is the volume within that region for flat space.  Let's now denote the mass in terms of the mass of the sun. Then in the LIGO event, the initial masses were 29 and 36 solar masses, and the final block hole had a mass of 62. Similarly, we will designate the Schwarzschild radii as multiples of the solar Schwarzschild radius of 3 km. We now estimate the change in the volume of space that takes place when two formerly distant black holes merge to be
\begin{equation}
\Delta V_{total}  =  \Delta V(R_1,R_2,62)  -  \Delta V(R_1,R_2,29)  -  \Delta V(R_1,R_2,36)
\end{equation}
representing the fact that the original black holes of mass 29 and 36 are gone (along with their volume excess) and are replace by one black hole with mass 62. 

We now estimate the change in volume that takes place during the emission of the gravitational wave in the near-in region in which the wave is generated. We take that to be at the location $R_2 = 4 R_s$ (corresponding to 0.37 sec on the LIGO time scale of Figs 1 \& 2) and at of the peak of the pulse at $R_1 = 1.5 R_s$ (corresponding to 0.42 sec on the LIGO time scale).  

Finally, we estimate the additional time created in the event as seconds. $\sqrt[\leftroot{-2}\uproot{2}3]{\Delta V_{total}}/c \approx 0.0012$ seconds.  With the additional gravitational redshift, this value will be larger, approximately 0.0015 sec = 1.5 ms. This is close to the value we obtained from dimensional analysis.  A more precise calculation for the volume change could be done using the detailed relativistic model used by the LIGO team to fit the pulse shape.

\section{Falsification/Verification}
\label{sec:FV}
A test of the \textit{now} theory might consist of a search for an unexpected (non-general relativity) lag of the peak of the signal, which takes place (seen in Fig 1) at 0.042 seconds, and to compare this arrival time to the time predicted by careful measurement of the pulse at earlier times (e.g. in the interval 0.34 to 0.36 seconds). We estimate, however, that the peak of the pulse seen in Figure 1 can be located to within a few milliseconds; moreover, the weaker parts of the preceding signal, where the 30 Hz oscillation is dominant, have even greater uncertainty.  Thus the one observed event falls short of providing a meaningful test of the theory. 

We have looked in detail at the raw LIGO signal, and tried applying several different digital filtering processes to see if we could obtain a time lag estimate better than we could get from the published LIGO pulse shape. We were not able to do so. 

There is a data analysis issue that must be addressed if we are to use a future LIGO event, one with improved signal-to-noise ratio, to test our prediction. The numerical model for the pulse has several adjustable parameters, in particular, the masses of the initial components and their separation, as well as various angles of the system with respect to the direction of the Earth.  If there is a small time delay that occurs, due to the creation of new space, then we are concerned that can easily be accommodated in the fit by a small adjustment in the size of the merging components or their distance from each other. For a true verification, all these parameters would be obtained exclusively from the earlier part of the pulse, and that would be used to predict the phase of the pulse peak; any deviation from that predicted phase could then be used to test our \textit{now} theory hypothesis.

\section{Modifying General Relativity}
\label{sec:MGR}
The calculations of the preceding sections were heuristic and not based on any specific reformulation of the general theory of relativity.  We make no claim in this paper to having completed such a new theory.  However it is worthwhile discussing ways of doing so.  In particular, how to assure that we will satisfy the correspondence principle, the need to preserving those aspects of general relativity that have already been extensively tested in the weak field realm.

One approach could be to leave the geometric aspects of general relativity untouched, but change the energy-momentum tensor, particularly for the vacuum. This is inspired, in part, by the method used to interpret the Hubble acceleration; the cosmological constant $\Lambda$ is absorbed into the energy-momentum tensor of the vacuum as \textit{dark energy}. For our purposes, we might want to add an additional term, perhaps one that consists of time derivatives  that would contribute significantly to the Einstein equations only when the spatial metric is changing rapidly with time; such a term would vanish in the static limit, and thereby carry over all the classical predictions of relativity theory (including the deflection of light and the precession of Mercury). 

We note that such a modification might not be discoverable in a homogeneous and isotropic expansion of the universe.  The reason is that any such universe is described by the Freedman-Walker metric:
\begin{equation}
ds^2 = -dt^2 + a(t) \left( \frac{dr^2}{1+kr^2/R^2} + r^2 d\Omega^2   \right)
\end{equation}
and any variation in the time coordinate could be absorbed into the expansion function a(t), and become indistinguishable from a change in the rate of the Hubble expansion.  It is conceivable that part of the observed dark energy actually consists of such a variation in the production of time, but cannot be distinguished, even in principle, unless we have an independently verified theory for that acceleration. In that case, a difference in the predicted vs observed redshift might be accounted for the by additional \textit{now} theory time lag.

If, however, there are inhomogeneous terms in the metric, then the effect might be detectable. Such is the situation in the early universe, when during a period of inflation (which, in our model, would include inflation in time as well as in space) when inhomogeneities are large and result in energy transfer into gravitational waves. A similar situation arises in merging black holes, as in the events reported by the LIGO team. 

One way to study this problem is to introduce time-dependent terms into the metric tensor for known static configurations, and to deduce what changes in the energy-momentum tensor would be needed to account for them.  We've done this with the Schwarzschild metric, introducing a term $[\partial g_{rr}(r,t) / \partial t]  dt^2$ that would not contribute to the static solutions, but would be important only in rapidly changing configurations, such as in the merging of two black holes. However, this approach has not yielded any result that seems plausible. 

With this approach, we are not proposing a change to general relativity, but simply a more complex vacuum, containing not only the dark energy, but an additional term that generates proper time when it is disturbed.  It is also conceivable that the term could be added to the energy-momentum tensor of matter, again playing a major role only when the distribution of mass-energy is rapidly changing with time.

\section{Discussion}
\label{sec:DIS}

A natural question arises: why are the new \textit{nows} created at the end of time, rather than uniformly throughout time, in the same way that new space is uniformly created throughout the universe. Why the asymmetry?  We have two possible ways of addressing this conundrum. 
 
The first possible resolution is that uniform creation of space is not, even in principle, observable. It is no more than a convenient and intuitive way of thinking about the Hubble expansion.  In the standard model it is a feature of the FLRW coordinate system. Instead, we could choose a static coordinate system, one in which there is no uniform expansion of space, but in which the Hubble expansion has the galaxies in motion. In this coordinate system the creation of new space takes place at the expanding event horizon at the edge of the universe.  These coordinates have both new space and new time created at the 4D periphery of the space-time. In this coordinate system the asymmetry vanishes, suggesting that it was only a coordinate issue, not a physical one.

A second possible explanation is that a physics principle of causality accounts for the apparent asymmetry in the creation of new space and new time.  In this view, we postulate that new time can be created only at the end of previous time, since its creation earlier would disrupt the causal connection of past events. Such a constraint does not occur for creation of space. Thus it is possible that causality allows the creation of new time only at the end of time. 

We find neither of these explanations to be totally satisfying, but that does not mean that either is false. The fact is that time does behave differently from space, so the unification of the two has always had an unexplained and unsatisfactory aspect. 

It is intriguing to consider whether reduction in space can result in the backwards flow of time.  The most philosophically intriguing example would be a collapsing universe. We can speculate that with no new \textit{nows}, that free will would not exist, and the universe would be vastly different from the one we experience. Unfortunately, this hypothesis seems to be completely untestable, since we have only one universe, one that is expanding, and we are apparently stuck with that one.

However, to result in a negative flow, the destruction of time of time would have to exceed the cosmological increase.  Note that in our merging black hole example, the creation of new space amounted to about 1 millisecond, over a time period (the duration of the event) of about 100 milliseconds.  Thus the newly created space, even in this incredibly violent event (the greatest local change in space-time ever observed) still was only a perturbation. In supernovae, the outward shell explosion is driven by the energy released from an even more violent core implosion driven by gravitational collapse, so the net creation of new space (and new time) dominates. 

Perhaps the most rapid dissipation of energy expected from theory takes place during black hole evaporation.  Would it cause local time reversal?  To do so, the local reversal rate would have to exceed the cosmological time increase rate. Using standard equations for black hole evaporation, we calculate that this reversal would take place only for black holes with mass less than $2.4\times 10^{10}$ kg, with Schwarzschild radius $10^{-25}$ meters.  This puts us deep into the realm of quantum physics. We note that in quantum field theory, very small, localized and rapid events contain amplitudes that can be interpreted as taking place in reversed time. Such reversed time is not, however, directly observable.  

Finally, we emphasize our hope that future analysis of LIGO events will not be aimed solely at verifying the general theory of relativity, but will allow us to detect unexpected departures from that theory. This can happen only if the fitting procedures are sufficiently overly constrained that a misfit cannot be compensated by parameter adjustment.

\acknowledgments

RM thanks Robert Rohde for help in analyzing the LIGO events. SM thanks John Preskill, Rana Adhikari and Sean Carroll for helpful conversations on related topics. 



\begin{thebibliography}{99}
\bibitem{integral}
The indefinite integral can be evaluated explicitly, and is given by: 
$\frac{1}{6}\pi r \left(8r^2 + 10r R_s^2 + 15 R_s^2 \right)  \sqrt{1 + R_s/r} + \frac{5}{4} \pi R_s^3 \log\left[ -R_s + 2r\left(1 + \sqrt{1-R_s/r}\right)\right]$
\end{thebibliography}
\end{document}